# Solo citations, duet citations, and prelude citations: New measures of the disruption of academic papers

Qiang Wu[1*]    Zhaoyang Yan[2]

[1]School of Management, University of Science and Technology of China, 96 Jinzhai Road, Hefei 230026, China.   [*Corresponding author. E-mail: qiangwu@ustc.edu.cn]
[2]Department of Computer Science and Technology, Shandong University, 180 Wenhua West Road, Weihai 264209, China.

**Abstract**   It is important to measure the disruption of academic papers. According to the characteristics of three different kinds of citations, this paper borrows musical vocabulary and names them solo citations (SC), duet citations (DC), and prelude citations (PC) respectively. Studying how to measure the disruption of a published work effectively, this study analyzes nine indicators and suggests a general evaluation formula. Seven of the nine indicators are innovations introduced by this paper: SC, SC-DC, SC-PC, SC-DC-PC, (SC-DC)/(SC+DC), (SC-PC)/(SC+DC), and (SC-DC-PC)/(SC+DC), as is the general formula. These indices are discussed considering two cases: One case concerns the Citation Indexes for Science and the other concerns Co-citations. The results show that, compared with other indicators, four indicators (SC, SC-DC, SC/(SC+DC), and (SC-DC)/(SC+DC)) are logically and empirically reasonable. Future research may consider combining these indices, for example, using SC multiplied by SC/(SC+DC) or SC-DC multiplied by (SC-DC)/(SC+DC), to get final evaluation results that contain desirable characteristics of two types of indicators. Confirming which of the evaluation results from these indicators can best reflect the innovation of research papers requires much empirical analysis.

**Introduction**

To measure the impact, value, or quality of academic papers, scholars have proposed many methods, such as the total citation number, h-index (Hirsch, 2005; Schubert, 2009), m-index (Bornmann, Mutz, & Daniel, 2008; Thor & Bornmann, 2011), method based on the Google's PageRank (Ma, Guan, & Zhao, 2008), citation-based summarization methods (Galgani, Compton, & Hoffmann, 2015), $fp^k$-index (Hu, Rousseau, & Chen, 2011), w-index (Wu, 2010; Yan, Wu, & Li, 2016), etc. Recently, Funk and Owen-Smith (2017) proposed a new method to measure consolidating and destabilizing technologies. Wu, Wang, and Evans (2019) used their methods to study the disruption of papers, patents, and software products and concluded that work by small teams is more disruptive than that by large teams, but there are known problems with this method (Wu & Wu, 2019). The research focus of the current paper is whether more effective evaluation methods can be designed.

**Measures of the Disruption of Academic Papers**

Wu, Wang, and Evans (2019) divided the number of citations into three categories, and constructed a basic formula for measuring disruptiveness, which has previously been put forward for patents (Funk & Owen-Smith, 2017). Their method of measuring the disruption of papers is as follows.

$$D=(SC-DC)/(SC+DC+PC) \qquad (1)$$

In Formula 1, D is the disruption of the focal paper, SC is the times that the other paper cites just the focal paper and does not cite its references, DC is the times that the other paper cites both the focal paper and any its references, and PC is the times that the other paper cites just any of the focal paper's references from the year after the focal paper is published. Here, the reason why this study chooses SC, DC, and PC to denote the three different types of citations is that according to their characteristics, this paper borrows musical vocabulary to name them solo citations (SC), duet citations (DC), and prelude citations (PC) respectively.

When different understandings of disruptiveness are used, different indexes are preferable to measure the disruption of a paper. Based on these three types of citations, there are at least other eight useful ways to measure the disruption of the focal paper as follows.

$$D=SC \quad (2)$$
$$D=SC-DC \quad (3)$$
$$D=SC-PC \quad (4)$$
$$D=SC-DC-PC \quad (5)$$
$$D=SC/(SC+DC) \quad (6)$$
$$D=(SC-DC)/(SC+DC) \quad (7)$$
$$D=(SC-PC)/(SC+DC) \quad (8)$$
$$D=(SC-DC-PC)/(SC+DC) \quad (9)$$

These indicators represent eight different perceptions. Except for Formula 6, which is slightly mentioned in Wu, Wang, and Evans' study (2019), the other seven indicators proposed by this paper are new methods for evaluating disruption. The characteristics of the nine indicators are summarized in Table 1.

TABLE 1.  Characteristics of the nine indicators

| Indicators | Condition | Viewpoint | | |
|---|---|---|---|---|
| | | SC | DC | PC |
| SC | | + | 0 | 0 |
| SC-DC | | + | - | 0 |
| SC-PC | | + | 0 | - |
| SC-DC-PC | | + | - | - |
| SC/(SC+DC) | | + | - | 0 |
| (SC-DC)/(SC+DC) | | + | - | 0 |
| (SC-PC)/(SC+DC) | SC > PC | + | - | - |
| | SC < PC | + | + | - |
| (SC-DC-PC)/(SC+DC) | SC > PC/2 | + | - | - |
| | SC < PC/2 | + | + | - |
| (SC-DC)/(SC+DC+PC) | SC > DC | + | - | - |
| | SC < DC | + | - | + |

*Note.* For the increase of the disruption, + plays a positive role; - plays a negative role; 0 does not play a role. All formulas should ensure that their denominator is not zero.

Total citations (TC=SC+DC), a very popular indicator, is a simple way to measure the impact of papers. At this time, the understanding of SC and DC is positive (+). The disruption calculated from Formulas 2, 3, 4, and 5 involves a quantity which should be correlated with the total number of citations. Dividing Formulas 2, 3, 4, and 5 by SC+DC yields Formulas 6, 7, 8, and 9, respectively. The disruption calculated by the last four formulas is a ratio. The advantage of these latter four formulas is that they can eliminate the differences caused by the number of citations (reflecting the popularity of each discipline or topic) among different disciplines and research topics. The disadvantage is that it is advantageous to papers with fewer citations and disadvantageous to papers with more citations. Therefore, when using such formulas, it is better to require a threshold for the number of citations, for example, that the paper must be cited more than 20 times before the ratio can be used.

Formulas 1, 8, and 9 have serious problems in practical application because of inconsistent understandings of DC or PC. For example, in the formula D=(SC-PC)/(SC+DC), when the numerator is greater than zero, DC is considered to play a positive role, and when it is less than zero, DC is considered to play a negative role. This inconsistency leads to the inability of these three indicators to be used directly, especially when the positive and negative data appear together.

**Empirical Analysis**

The two cases chosen for this study are the groundbreaking achievements related to "Citation Indexes for Science" (Garfield, 1955) and "Co-citation" (Small, 1973). These aspects are popular and important in the field of Library and Information Science. Their importance can be demonstrated by the increasing number of annual citations of such papers, especially since 2005 (see Figures 1 and 2). The citation data were obtained from the Web of Science Core Collection database on April 9, 2019.

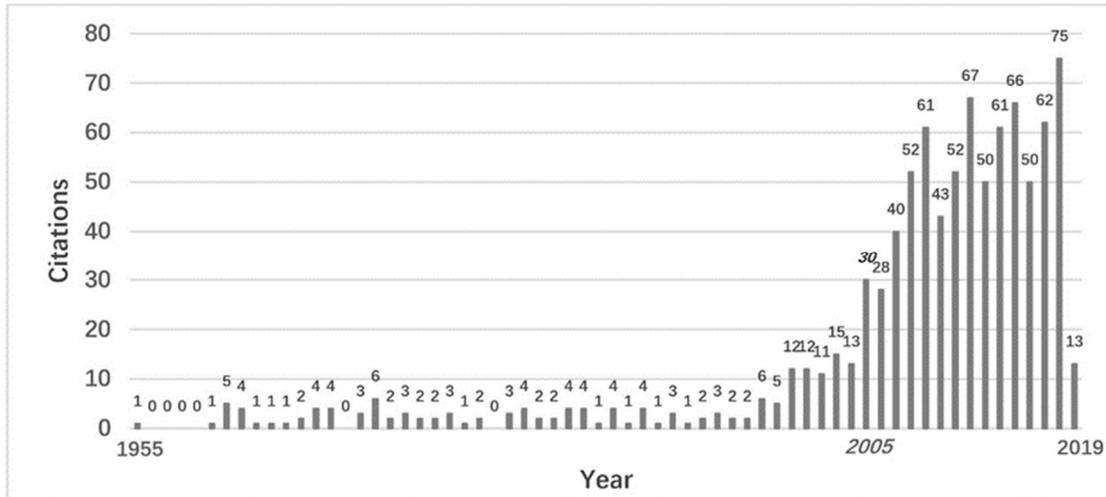

FIG. 1.   Number of annual citations of papers on "Citation Indexes for Science" (Garfield, 1955) (1955-2019)

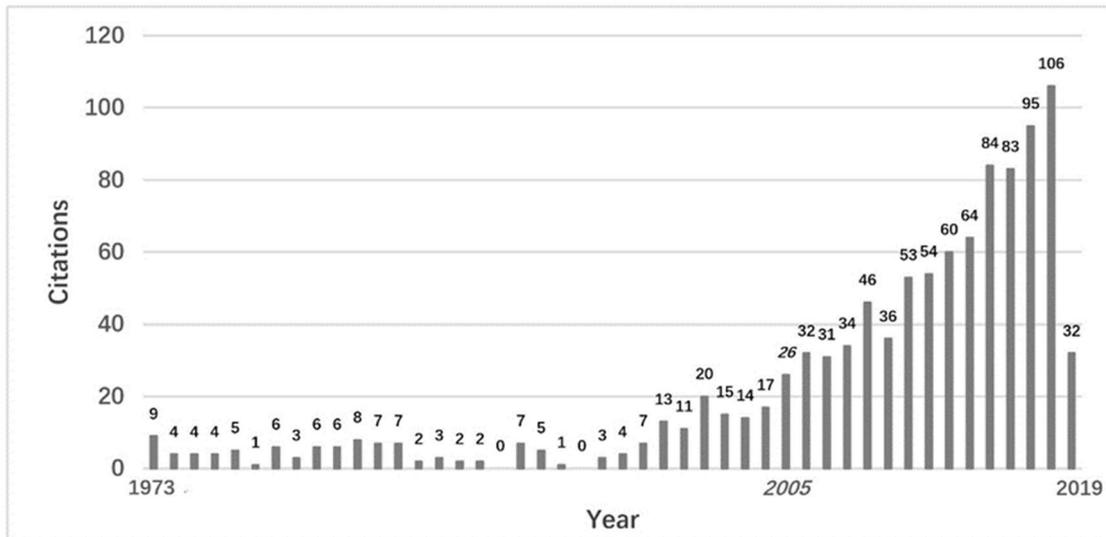

FIG. 2.   Number of annual citations of papers on "Co-citation" (Small, 1973) (1973-2019)

TABLE 2.   Scores of the nine indicators in Cases I and II

| Indicators | Case I | Case II |
|---|---|---|
| SC | 843 | 647 |
| SC-DC | 771 | 262 |
| SC-PC | -388 | -6054 |
| SC-DC-PC | -460 | -6439 |
| SC/(SC+DC) | 0.9213 | 0.6269 |
| (SC-DC)/(SC+DC) | 0.8426 | 0.2539 |
| (SC-PC)/(SC+DC) | -0.4240 | -5.8663 |
| (SC-DC-PC)/(SC+DC) | -0.5027 | -6.2393 |
| (SC-DC)/(SC+DC+PC) | 0.3593 | 0.0339 |

From these two examples, it is not difficult to see that because the values of SC-PC, SC-DC-PC, (SC-PC)/(SC+DC), and (SC-DC-PC)/(SC+DC) are negative, they cannot reflect the innovation of these two achievements (i.e., "Citation Indexes for Science" and "Co-citation"), so these are inappropriate as evaluation indices. The main reasons are that PC is too large and that most of these papers and the focal paper belong to different disciplines or research topics. Therefore, when measuring disruption with formulas containing PC, caution must be exercised. Only when the focal paper and these papers are relatively close to the same research topics can the method be effective. As for (SC-DC)/(SC+DC+PC), although all of the indicator's scores are positive, for Case II, the value is only 0.0339, which is too small. Compared with 0.3593 in Case I, the difference between the values is too big, which means the indicator is unreasonable. Besides, we can see from Table 1 that there is a serious defect in this indicator, that is, the perception of PC is inconsistent (Wu & Wu, 2019). For the other four indicators (i.e., SC, SC-DC, SC/(SC+DC), and (SC-DC)/(SC+DC)), no obvious defects have been found in the analysis of their characteristics considering these two cases. Of course, when using SC/(SC+DC) and (SC-DC)/(SC+DC), we must pay attention to the influence of total citations: it is beneficial to the papers with fewer citations, and it is unfavorable to the papers with more citations. Therefore, perhaps a more reasonable way is to mix these indices for disruption measurement.

We now put forward a feasible, general method to combine SC, SC-DC, SC/(SC+DC), and (SC-DC)/(SC+DC) for comprehensive evaluation. The general evaluation formula is as follows.

$$D=(SC^a-DC^b)^c/(SC^d+DC^e)^f \qquad (10)$$

where a, b, c, d, e, and f are coefficients which can be positive, negative, or 0. The formula should ensure that its denominator is not zero.

When these coefficients take on certain specific values, Formula 10 will degenerate into other formulas, as follows:

a=1, b=0, c=1, and f=0,           D=SC

a=1, b=1, c=1, and f=0,           D=SC-DC

a=1, b=0, c=1, d=1, e=1, and f=1,    D=SC/(SC+DC)

a=1, b=1, c=1, d=1, e=1, and f=1,    D=(SC-DC)/(SC+DC)

a=1, b=0, c=2, d=1, e=1, and f=1,    D=SC$^2$/(SC+DC)=SC*SC/(SC+DC)

a=1, b=1, c=2, d=1, e=1, and f=1,    D=(SC-DC)$^2$/(SC+DC)= (SC-DC)* (SC-DC)/(SC+DC)

The latter two formulas are essentially ways of multiplying SC by SC/(SC+DC), or (SC-DC) by (SC-DC)/(SC+DC). In fact, multiplying SC by (SC-DC)/(SC+DC) can be used as another new indicator. Much empirical analysis is needed to confirm which of the evaluation results of these indicators can best reflect the innovation of a paper.

**Conclusion and Discussion**

With the goal of measuring the disruption of a paper effectively, this study analyzes the characteristics of nine indicators, among which seven are new: SC, SC-DC, SC-PC, SC-DC-PC, (SC-DC)/(SC+DC), (SC-PC)/(SC+DC), and (SC-DC-PC)/(SC+DC)). A general evaluation formula is also introduced to generate other potentially useful indicators. Then, these indices are discussed considering two aspects which are popular research topics in this area: "Citation Indexes for Science" and "Co-citations". The results show that, compared with other indicators, four of the indicators (i.e., SC, SC-DC, SC/(SC+DC), and (SC-DC)/(SC+DC)) are logically and empirically reasonable. In future research, a more useful indicator may combine these indices, for example, SC-DC multiplied by (SC-DC)/(SC+DC) or SC multiplied by SC/(SC+DC), to obtain final evaluation results. Such combination indices could embody the best characteristics of two kinds of indicators.

To measure the disruption of academic papers from multiple angles, more variables can be added to the formulas mentioned above. For example, when scholars hold a negative (-) attitude towards the reference of the focal paper, the number of references (NR) can be introduced into the formula and SC-NR, SC-DC-NR, SC-PC-NR and SC-DC-PC-NR can be obtained. Moreover, when researchers have a positive (+) attitude towards DC, we can get formulas as follows: SC+DC-NR, SC+DC-PC, SC+DC-PC-NR, etc. All of these formulas can be used to evaluate the disruption. Kosmulski (2011) has used the formula SC+DC-NR > 0 to define the concept of "successful papers". NR can also be placed in the denominator. Yanovsky (1981) has proposed using (SC+DC)/NR to access the impact of scientific journals. Although the research object of this study only focuses on papers, the various methods proposed in this research can also be used to evaluate patents, journals, software products, web pages and other objects with links that are similar to citations between papers.

**Acknowledgements**

This work was supported in part by the National Natural Science Foundation of China (grant number 71874173) and the Academic Division of Mathematics and Physics of the Chinese Academy of Sciences (grant number 2018-M04-B-004).

**References**

Bornmann, L., Mutz, R., & Daniel, H. D. (2008). Are there better indices for evaluation purposes than the h index? A comparison of nine different variants of the h index using data from biomedicine. Journal of the American Society for Information Science and Technology, 59(5), 830-837.

Funk, R. J., & Owen-Smith, J. (2017). A dynamic network measure of technological change. Management Science, 63(3), 791-817.

Galgani, F., Compton, P., & Hoffmann, A. (2015). Summarization based on bi-directional citation analysis. Information Processing and Management, 51(1), 1-24.

Garfield, G. (1955). Citation indexes for science: A new dimension in documentation through association of ideas. Science, 122(3159), 108-111.

Hirsch, J. E. (2005). An index to quantify an individual's scientific research output. Proceedings of the National Academy of Sciences of the United States of America, 102(46), 16569-16572.

Hu, X., Rousseau, R., & Chen, J. (2011). On the definition of forward and backward citation generations. Journal of Informetrics, 5(1), 27-36.

Kosmulski, M. (2011). Successful papers: A new idea in evaluation of scientific output. Journal of Informetrics, 5(3), 481-485.

Ma, N., Guan, J., & Zhao, Y. (2008). Bringing PageRank to the citation analysis. Information Processing and Management, 44(2), 800-810.

Schubert, A. (2009). Using the h-index for assessing single publications. Scientometrics, 78(3), 559-565.

Small, H. (1973). Co-citation in scientific literature: A new measure of the relationship between two documents. Journal of the American Society for Information Science, 24(4), 265-269.

Thor, A., & Bornmann, L. (2011). The calculation of the single publication hindex and related performance measures. Online Information Review, 35(2), 291-300.

Wu, L., Wang, D., & Evans, J. A. (2019). Large teams develop and small teams disrupt science and technology. Nature, 566(7744), 378-382.

Wu, Q. (2010). The w-index: A measure to assess scientific impact by focusing on widely cited papers. Journal of the American Society for Information Science and Technology, 61(3), 609-614.

Wu, S., & Wu, Q. (2019, April 19). A confusing definition of disruption. https://doi.org/10.31235/osf.io/d3wpk

Yan, Z., Wu, Q., & Li, X. (2016). Do Hirsch-type indices behave the same in assessing single publications? An empirical study of 29 bibliometric indicators. Scientometrics, 109(3), 1815-1833.

Yanovsky, V. I. (1981). Citations analysis significance of scientific journals. Scientometries, 3(3), 223-233.